\documentclass[12pt]{article}
\usepackage{amsfonts,amsmath}
\usepackage{graphicx}
\usepackage{amsmath,amssymb,fancyhdr}

\def\firstpage{\hfill RUP-13-16}
\makeatletter
\def\ps@titlepage{%
	\@oddhead{\hfil\firstpage\hfil}%
}
\makeatother

\makeatletter

\@addtoreset{equation}{section}
\makeatother

\newcommand{\be}{\begin{eqnarray}}
\newcommand{\ee}{\end{eqnarray}}
\newcommand{\nn}{\nonumber}
\newcommand{\nl}{\nonumber \\}
\newcommand{\pd}{\partial}

\title{{\bfseries Non-$\gamma_{5}$hermiticity fermions in two dimensions}}
\author{Syo Kamata and Hidekazu Tanaka\\
{\it Department of Physics, Rikkyo University, Tokyo 171-8501, Japan}\\
{\it E-mail:} {\ttfamily skamata@rikkyo.ac.jp, tanakah@rikkyo.ac.jp}}
\date{}

\begin{document}
\maketitle
\thispagestyle{titlepage}

\begin{center}
\Large{Abstract}
\end{center}

We construct 2D non-$\gamma_{5}$hermiticity fermions based on the minimal doubling fermion.
We investigate symmetries, reflection positivity, eigenvalue distribution and the number of poles for our fermions.
As simple tests for application to the fermion, the Gross-Neveu model in two dimensions is studied using the non-$\gamma_{5}$hermiticity fermion.
We draw the parity broken phase diagram, called Aoki phase and the chiral broken phase diagram for the model with an imaginary chemical potential.

\section{Introduction}
The lattice gauge theory is a useful tool for investigating strong coupling physics \cite{Wilson:1974sk}.
As well known, lattice fermions have a serious problem, called the doubling problem.
The naive fermion has $2^{d}$ degenerate spectra, called doublers which are eight pairs of $\pm$-chiral charge, and the spectra directly affect physics, e.g. asymptotic free, chiral anomaly and so on.
Hence, the problem makes it difficult to reveal the physics, including fermions.

Unfortunately, a no-go theorem was discovered by Nielsen and Ninomiya \cite{Nielsen:1980rz}-\cite{ Nielsen:1981hk} which states that we cannot construct a single pole fermion preserving all of the symmetries or properties: translation invariance, chiral symmetry, locality and ($\gamma_{5}$)hermiticity.
To overcome this problem, many fermions have been formulated, e.g., the Wilson fermion, which breaks chiral symmetry \cite{Wilson:1974sk}, the KS fermion, which regards doublers as flavors \cite{Kogut:1974ag}, the SLAC fermion, which does not preserve locality\cite{Drell:1976mj} and so on.
In recent years, minimal doubling fermions with only two doublers have been formulated by Karsten et al. \cite{Karsten:1981gd}-\cite{ Borici:2007kz}.
The minimal doubling fermion preserves exact chiral symmetry but breaks some discrete symmetries and cubic symmetry.
We expect that minimal doubling fermion is a new tool for investigating QCD physics.

Finite temperature and density physics are main subjects for the lattice gauge theory.
In a finite density theory, it is well known that a fermion bilinear term in the action is broken $\gamma_{5}$hermiticity by a chemical potential term.
The $\gamma_{5}$hermiticity guarantees the hermiticity of the Hamiltonian and is also a reality condition for fermion determinant appearing when fermions are integrated out from the partition function.
In general, the fermion determinant in a finite density theory is not a real number but rather a complex number.
For the estimation of observables, we need to use an appropriate reweighting method.
However, in the high density region, the complex phase of fermion determinant fluctuates in a wide range, therefore, the observable approaches to zero and can hardly be estimated.
No one knows general resolutions for this problem, and this problem is still an open problem, called the sign problem.

In this paper, we formulate 2D fermions without $\gamma_{5}$hermiticity (non-$\gamma_{5}$hermiticity fermions) based on the minimal doubling fermion.
As with the minimal doubling fermion, the non-$\gamma_{5}$hermiticity fermion breaks some discrete symmetries.
We obtain the eigenvalue distribution and the number of poles for the fermions and discuss the selection rule for an optimum fermion to apply to a practical analysis.
For simple application tests, the 2D Gross-Neveu model is studied using the non-$\gamma_{5}$hermiticity fermion.
We draw two sorts of phase diagrams, parity broken phase diagrams, called Aoki phase and chiral broken phase diagrams in massless and an imaginary chemical potential system.
By the analyzing these models with the fermion, we expect that we can more deeply understand the structure of lattice fermions, and thus the sign problem.

This paper is organized as follows.
In Sect. \ref{sec:non-gamma5}, we construct non-$\gamma_{5}$hermiticity fermions based on the minimal doubling fermion and investigate their symmetries and properties.
In Sect. \ref{sec:aoki}, we study parity broken phase diagrams for the 2D Gross-Neveu model using the non-$\gamma_{5}$hermiticity fermion.
In Sect. \ref{sec:ichem}, we also draw chiral broken phase diagrams for the Gross-Neveu model adding a imaginary chemical potential in two dimensions.
In Sect. \ref{sec:real-cond}, we discuss a reality condition for observables from the eigenvalue distribution of the fermions with an imaginary chemical potential. 
The Final section is devoted to the summary.

\section{2D non-$\gamma_{5}$hermiticity fermions} \label{sec:non-gamma5}
In this section, we define 2D fermions without $\gamma_{5}$hermiticity (non-$\gamma_{5}$hermiticity fermions) based on the minimal doubling fermion and investigate the symmetries and properties of the fermions.
The minimal doubling fermions were formulated by Karsten et al. \cite{Karsten:1981gd}-\cite{ Borici:2007kz} and do not interfere with the no-go theorem of Nielsen-Ninomiya because two doublers, which are a $\pm$-chiral charge pair, appear.
The fermions preserve translation invariance, chiral symmetry, locality and $\gamma_{5}$hermiticity, but break (hyper)cubic symmetry and some discrete symmetries, e.g. charge conjugation, parity symmetry and so on.
We refer the reader to Refs. \cite{Karsten:1981gd}-\cite{Kamata:2011jn} for more details on the minimal doubling fermion in detail.

Now, we construct non-$\gamma_{5}$hermiticity fermions adding PT symmetry breaking terms because lattice fermions with PT-symmetric doubler-suppressing kinetic terms always generate $2^{d}$ doublers \cite{Kamata:2011jn}.
We define five free massless non-$\gamma_{5}$hermiticity fermions in coordinate space as follows (lattice spacing $a=1$)
\footnote{We define the non-$\gamma_{5}$hermiticity fermions up to exchanging a temporal index and a spatial one: In the $D^{(2)}_{1}$ case, $\bar{D}_{1}^{(2)}(p_{1},p_{4}) =  \sum_{\mu=1,4} i \sin p_{\mu} \cdot \gamma_{\mu} + \kappa ( 1-\cos p_{1}  ) \cdot \gamma_{1}$. The $D^{(2)}_{1}$ preserves P symmetry but breaks T symmetry defined below.}:
\be
S &=& \sum_{n,m} \bar{\psi}_{n} D^{(2)}_{nm} \psi_{m},\nl
D^{(2)}_{1 \ nm} &=& \sum_{\mu=1,4} \frac{1}{2} \left( \delta_{n+\hat{\mu},m} - \delta_{n-\hat{\mu},m} \right) \cdot \gamma_{\mu} +\frac{\kappa}{2} \left(2 \delta_{n,m}-\delta_{n+\hat{1},m} - \delta_{n-\hat{1},m} \right) \cdot \gamma_{1}, \label{eq:mod2-1}\\
D^{(2)}_{2 \ nm} &=& \sum_{\mu=1,4} \frac{1}{2} \left( \delta_{n+\hat{\mu},m} - \delta_{n-\hat{\mu},m} \right) \cdot \gamma_{\mu} +\frac{\kappa}{2} \left(2 \delta_{n,m}-\delta_{n+\hat{4},m} - \delta_{n-\hat{4},m} \right) \cdot \gamma_{1}, \label{eq:mod2-2}\\
D^{(2)}_{3 \ nm} &=& \sum_{\mu=1,4} \frac{1}{2} \left( \delta_{n+\hat{\mu},m} - \delta_{n-\hat{\mu},m} \right) \cdot \gamma_{\mu} +\frac{\kappa}{2} \sum_{\mu=1,4} \left(2 \delta_{n,m}-\delta_{n+\hat{\mu},m} - \delta_{n-\hat{\mu},m} \right) \cdot \gamma_{\mu} \label{eq:mod2-3},\\
D^{(2)}_{4 \ nm} &=& \sum_{\mu=1,4} \frac{1}{2} \left( \delta_{n+\hat{\mu},m} - \delta_{n-\hat{\mu},m} \right) \cdot \gamma_{\mu} +\frac{\kappa}{2} \sum_{\mu,\nu=1,4 \ \mu \ne \nu} \left(2 \delta_{n,m}-\delta_{n+\hat{\mu},m} - \delta_{n-\hat{\mu},m} \right) \cdot \gamma_{\nu} \label{eq:mod2-4},\nl
\ee
\be
D^{(2)}_{5 \ nm} &=& \sum_{\mu=1,4} \frac{1}{2} \left( \delta_{n+\hat{\mu},m} - \delta_{n-\hat{\mu},m} \right) \cdot \gamma_{\mu} +\frac{\kappa}{2} \sum_{\mu=1,4} \left(2 \delta_{n,m}-\delta_{n+\hat{\mu},m} - \delta_{n-\hat{\mu},m} \right) \cdot \gamma_{1} \label{eq:mod2-5},\nl
\ee
and in momentum space,
\be
S &=& \int \frac{d^{2}p}{(2 \pi)^{2}} \ \bar{\psi}(-p) D^{(2)}(p) \psi(p), \nl
D_{1}^{(2)}(p) &=&  \sum_{\mu=1,4} i \ \sin p_{\mu} \cdot \gamma_{\mu} + \kappa ( 1-\cos p_{1}  ) \cdot \gamma_{1} ,\label{eq:mod2-1p}\\
D_{2}^{(2)}(p) &=&  \sum_{\mu=1,4} i \ \sin p_{\mu} \cdot \gamma_{\mu} + \kappa ( 1-\cos p_{4}  ) \cdot \gamma_{1} ,\label{eq:mod2-2p}\\
D_{3}^{(2)}(p) &=& \sum_{\mu=1,4}  i \ \sin p_{\mu} \cdot \gamma_{\mu}  + \kappa \sum_{\mu=1,4}( 1-\cos p_{\mu} ) \cdot \gamma_{\mu},\\
D_{4}^{(2)}(p) &=& \sum_{\mu=1,4}  i \ \sin p_{\mu} \cdot \gamma_{\mu}  + \kappa \sum_{\mu,\nu=1,4 \ \mu \ne \nu}( 1-\cos p_{\mu} ) \cdot \gamma_{\nu},\\
D_{5}^{(2)}(p) &=& \sum_{\mu=1,4}  i \ \sin p_{\mu} \cdot \gamma_{\mu}  + \kappa \sum_{\mu=1,4}( 1-\cos p_{\mu} ) \cdot \gamma_{1},
\ee
where $\kappa$ is a hopping parameter which is a real number
\footnote{We regard indices 1 and 4 as temporal and spacial directions, respectively.}. 
Note that $D^{(2)}_{3}$ at $\kappa=-1(+1)$ is just a Dirac operator with a forward(backward) difference operator. 
Next, we define charge conjugation(C), parity transformation(P) and time reversal(T) as follows:
\be
\mathrm{C} &:& \ D(p_{1},p_{4}) \rightarrow  -C D^{\top} (-p_{1},-p_{4}) C^{-1},\nl
\mathrm{P} &:& \ D(p_{1},p_{4}) \rightarrow \gamma_{4} D(-p_{1},p_{4}) \gamma_{4},\\
\mathrm{T} &:& \ D(p_{1},p_{4}) \rightarrow \gamma_{1} D(p_{1},-p_{4}) \gamma_{1},\nn
\ee
where $C$ is a charge conjugation matrix.
In two dimensions, we can define $C=i \gamma_{1}$ as a charge conjugation matrix where $\gamma_{1}=\sigma^{2}$ and $\gamma_{4}=\sigma^{1}$.
We also define a Dirac operator preserving chiral symmetry and $\gamma_{5}$hermiticity as follows:
\be
\mathrm{chiral} &:& \ D(p_{1},p_{4})=-\gamma_{5} D(p_{1},p_{4}) \gamma_{5}, \nl
\mathrm{\gamma_{5}\,hermiticity} &:& \ D(p_{1},p_{4})=\gamma_{5} D^{\dagger}(p_{1},p_{4}) \gamma_{5}, \nn
\ee
where $\gamma_{5}$ is a chiral matrix defined as $\gamma_{5}=i \gamma_{1} \gamma_{4}$.
\begin{table}[t]
  \begin{center} 
  \begin{tabular}{|c||c|c|c|c|c|c|c||c|c|}\hline
 & C & P & T & CP & CT & PT & CPT  & chi & $\gamma_{5}$h \\ \hline \hline
$D_{1}^{(2)}$ & $\times$ & $\times$ & $\bigcirc$ & $\bigcirc$ & $\times$ & $\times$ & $\bigcirc$ & $\bigcirc$ & $\times$  \\\hline 
$D_{2}^{(2)}$ & $\times$ & $\times$ & $\bigcirc$ & $\bigcirc$ & $\times$ & $\times$ & $\bigcirc$ & $\bigcirc$ & $\times$   \\\hline 
$D_{3}^{(2)}$ & $\times$ & $\times$ & $\times$ & $\times$ & $\times$ & $\times$ & $\bigcirc$ & $\bigcirc$ & $\times$   \\\hline 
$D_{4}^{(2)}$ & $\times$ & $\times$ & $\times$ & $\times$ & $\times$ & $\times$ & $\bigcirc$ & $\bigcirc$ & $\times$   \\\hline 
$D_{5}^{(2)}$ & $\times$ & $\times$ & $\bigcirc$ & $\bigcirc$ & $\times$ & $\times$ & $\bigcirc$ & $\bigcirc$ & $\times$  \\\hline 
  \end{tabular}
  \end{center}
  \caption{The symmetric properties of discrete symmetries, chiral symmetry and $\gamma_{5}$hermiticity for the Dirac operators $D^{(2)}_{1}-_{5}$.}
  \label{tab:discrete}
\end{table}
All the Dirac operators have preserved chiral symmetry and CPT symmetry but broken C, P, CT and PT symmetries.
$D^{(2)}_{1}$, $D^{(2)}_{2}$, and $D^{(2)}_{5}$ also preserve T and CP symmetry, but $D^{(2)}_{3}$ and $D^{(2)}_{4}$ do not.
$D^{(2)}_{1}$, $D^{(2)}_{2}$ and $D^{(2)}_{5}$ have the discrete symmetries as the Karsten-Wilczek minimal doubling fermions \cite{Karsten:1981gd, Wilczek:1987kw}, and $D^{(2)}_{3}$ and $D^{(2)}_{4}$ have the same ones as the Borici-Creutz ones \cite{ Creutz:2007af, Borici:2007kz}.
We summarize their symmetric properties in Table \ref{tab:discrete}.

\begin{figure}[htbp]
\begin{minipage}{0.5\hsize}
\begin{center}
\includegraphics[width=70mm]{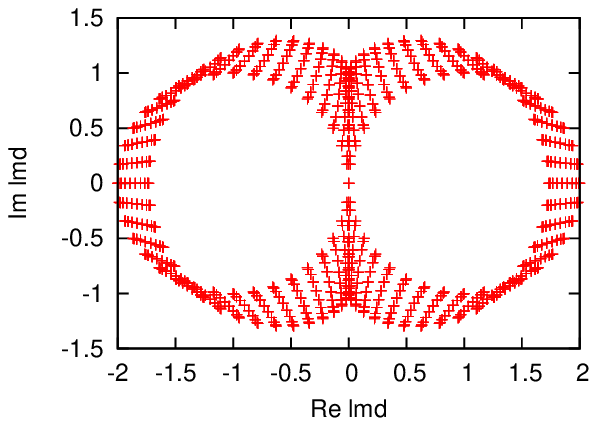}
\hspace{1.6cm}  (a)
\end{center}
\end{minipage}
\begin{minipage}{0.5\hsize}
\begin{center}
\includegraphics[width=70mm]{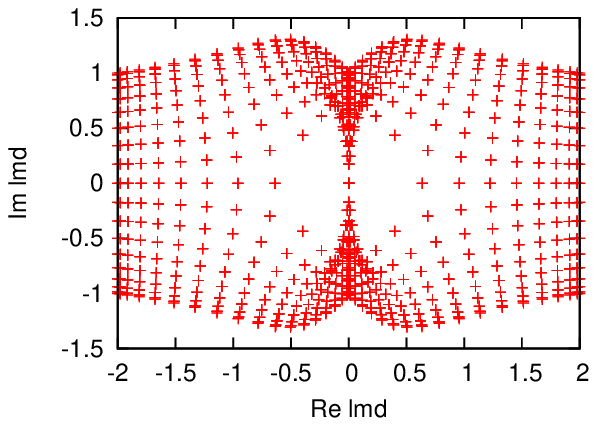}
\hspace{1.6cm}  (b)
\end{center}
\end{minipage}
\begin{minipage}{0.5\hsize}
\begin{center}
\includegraphics[width=70mm]{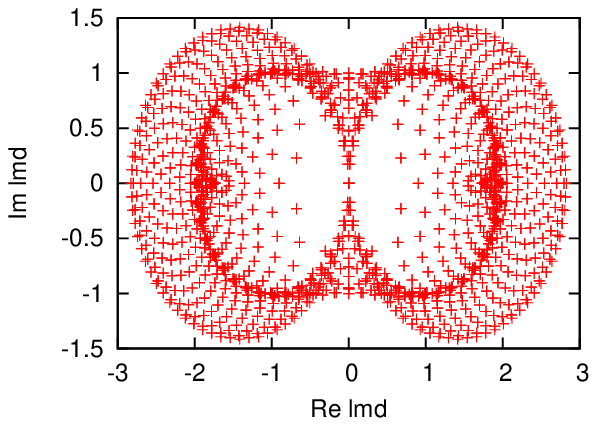}
\hspace{1.6cm}  (c)
\end{center}
\end{minipage}
\begin{minipage}{0.5\hsize}
\begin{center}
\includegraphics[width=70mm]{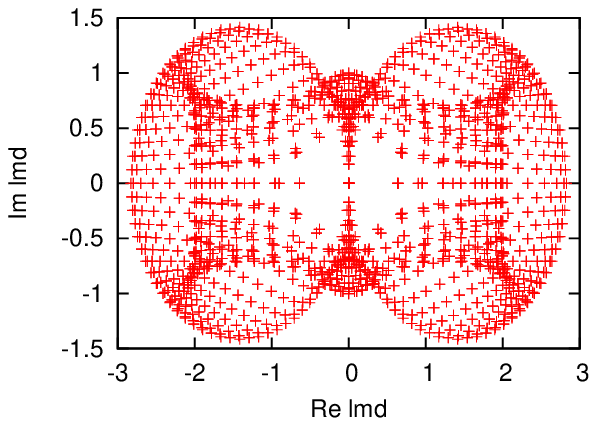}
\hspace{1.6cm}  (d)
\end{center}
\end{minipage}
\begin{minipage}{0.5\hsize}
\begin{center}
\includegraphics[width=70mm]{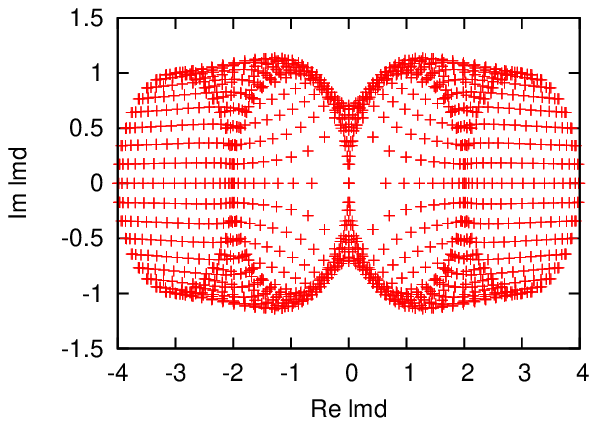}
\hspace{1.6cm}  (e)
\end{center}
\end{minipage}
\caption{The eigenvalue distribution of Dirac operators (a) $D^{(2)}_{1}$, (b) $D^{(2)}_{2}$, (c) $D^{(2)}_{3}$, (d) $D^{(2)}_{4}$ and (e) $D^{(2)}_{5}$. The horizontal and vertical axes denote the real and the imaginary parts of eigenvalues, respectively. The hopping parameter, the fermion mass and lattice size are fixed at $\kappa=1$, $m=0$, and $L^{2}=36 \times 36$. The blue circle points denote eigenvalues for momenta $p=(0,0)$, $(0,\pi)$, $(\pi,0)$, and $(\pi,\pi)$.}
\label{fig:eigenvalue}
\end{figure}

\begin{figure}[htbp]
\begin{minipage}{0.5\hsize}
\begin{center}
\includegraphics[width=70mm,angle=0]{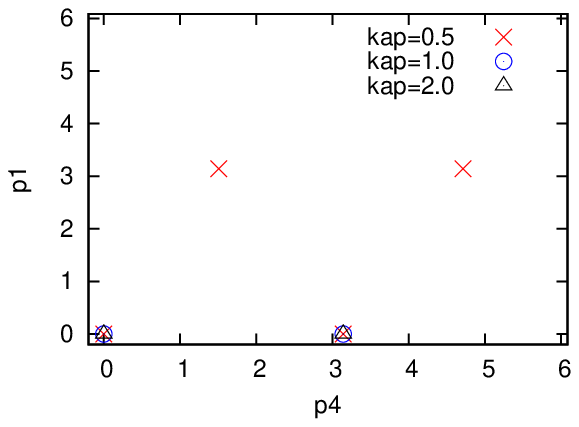}
\hspace{1.6cm}  (a)
\end{center}
\end{minipage}
\begin{minipage}{0.5\hsize}
\begin{center}
\includegraphics[width=70mm,angle=0]{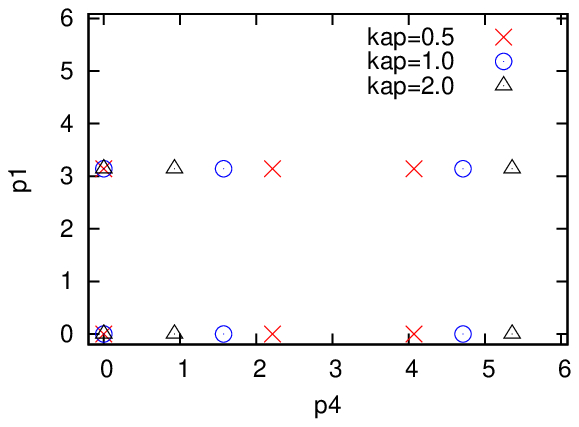}
\hspace{1.6cm}  (b)
\end{center}
\end{minipage}
\begin{minipage}{0.5\hsize}
\begin{center}
\includegraphics[width=70mm,angle=0]{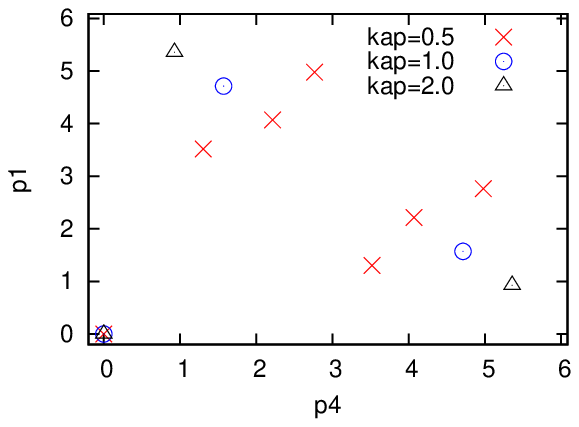}
\hspace{1.6cm}  (c)
\end{center}
\end{minipage}
\begin{minipage}{0.5\hsize}
\begin{center}
\includegraphics[width=70mm,angle=0]{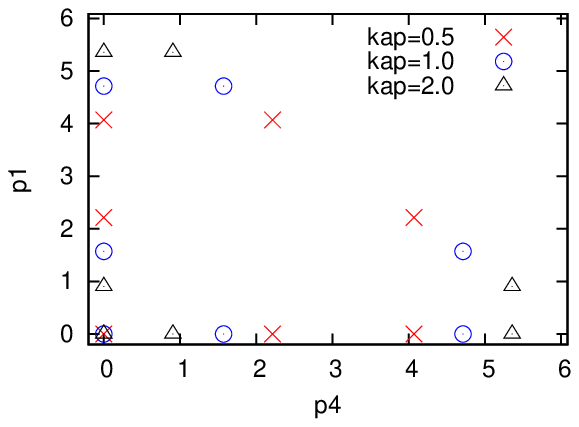}
\hspace{1.6cm}  (d)
\end{center}
\end{minipage}
\begin{minipage}{0.5\hsize}
\begin{center}
\includegraphics[width=70mm,angle=0]{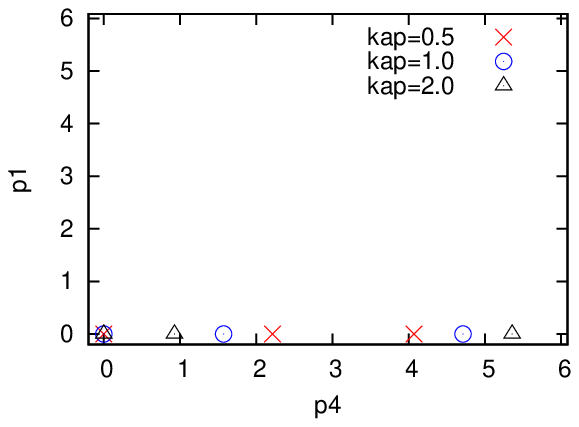}
\hspace{1.6cm}  (e)
\end{center}
\end{minipage}
\caption{The pole distribution of five Dirac operators (a) $D^{(2)}_{1}$, (b) $D^{(2)}_{2}$, (c) $D^{(2)}_{3}$, (d) $D^{(2)}_{4}$, and (e) $D^{(2)}_{5}$. The horizontal and vertical axes denote $p_{4}$ and $p_{1}$ respectively. The cross points, circle points and triangle points represent poles at $\kappa=0.5, 1$, and $2$, respectively.}
\label{fig:pole}
\end{figure}

To find the model application possibilities, we obtain the eigenvalue distribution and the number of doublers of the fermions.
Firstly, we present eigenvalue distribution in Fig.\ref{fig:eigenvalue}.
We clearly see that the eigenvalues of $D^{(2)}_{2}$, $D^{(2)}_{3}$, $D^{(2)}_{4}$ and $D^{(2)}_{5}$ spread around the origin.
The eigenvalues spread over the $\mathrm{Re}\lambda$-$\mathrm{Im}\lambda$ plane entirely within the continuum limit.
This is a typical feature of the sign problem.
In the $D^{(2)}_{1}$ case, by contrast, there are spaces that are enclosed eigenvalues in the plane.
In the continuum limit, the eigenvalues tend to the infinity boundary in the plane and are distributed along the $\mathrm{Im}\lambda$ axis.
The distribution of $D^{(2)}_{1}$ in the limit has the same features as the Dirac operators that satisfy $\gamma_{5}$hermiticity, e.g. the naive fermion.
In Fig.\ref{fig:pole}, we also present the number of poles for the fermions at $\kappa=0.5,1$, and $2$.
The figure shows that there are two poles, $p=(0,0)$ and $(0,\pi)$, in $D^{(2)}_{1}$ at $\kappa \ne 1$.
However, the other fermions have more than two poles.

As noted above, the eigenvalue distribution of $D^{(2)}_{1}$ in the continuum limit is identical with the continuum fermion, whose eigenvalues are distributed along the imaginary axis.
On the other hand, the eigenvalues of the others are distributed on the ${\rm Re}\lambda$-${\rm Im}\lambda$ plane entirely within the limit.
According to the lattice theory principles, a lattice fermion in the continuum limit should recover the continuum fermion, $\bar{\psi}ip\cdot \gamma \psi$.
Using appropriate regularization, e.g. small mass or anti-boundary condition, we can probably use $D^{(2)}_{2-5}$ to analyze a model.
By the lattice principle, however, the Dirac operators $D^{(2)}_{2-5}$ should be excluded from the candidates for application to a practical calculation.

According to the analysis of the number of poles, there are generally an odd number or more than four poles.
The instinctive reason for odd or many doublers is that the dispersion relations of the fermions are complex.
For example, in the $D^{(2)}_{2}$ case, the dispersion relation is derived as follows:
\be
\sum_{\mu=1,4} \sin^{2}p_{\mu} - \kappa^{2}(1-\cos p_{4})^{2}-2i \kappa \sin p_{1}(1-\cos p_{4})=0.\label{eq:dispersion}
\ee
The first term on the l.h.s. has an opposite signature to the second term.
This can easily lead to six real solutions for Eq. (\ref{eq:dispersion}),
\be
(p_{1},p_{4}) &=& (0,0),(\pi,0),\nl
             & &(0,4\tan^{-1}[\kappa+\sqrt{1+\kappa^{2}}]), (\pi,4\tan^{-1}[\kappa+\sqrt{1+\kappa^{2}}]),\nl
             & & (0,-4\tan^{-1}[\kappa-\sqrt{1+\kappa^{2}}]), (\pi,-4\tan^{-1}[\kappa-\sqrt{1+\kappa^{2}}]). \nl
\ee
In a similar way, the solutions for $D^{(2)}_{3}$ at $\kappa=-1$, which is the forward difference fermion, are obtained as
\be
(p_{1},p_{4}) &=& (0,0),(\pi/2,-\pi/2),(-\pi/2,\pi/2). \label{eq:D3pole}
\ee

Now, we study the relationship between doublers and $\gamma_{5}$hermiticity in detail.
We define a Dirac operator $D(p)$ which is assumed by continuity, periodicity, translation invariance, locality, and  $\gamma_{\mu}$-linear as follows:
\be
D(p) &=& \sum_{\mu,\nu,n_{\mu \nu}}  \left[ A_{\mu \nu n_{\mu \nu}} \sin(n_{\mu \nu} p_{\mu}) + B_{\mu \nu n_{\mu \nu}} \cos(n_{\mu \nu}p_{\mu}) \right] \cdot \gamma_{\nu} \nl
 &\equiv& \sum_{\nu} {\cal D}_{\nu}(p)  \cdot \gamma_{\nu},
\ee
where $A_{\mu \nu n_{\mu \nu}},B_{\mu \nu n_{\mu \nu}}$ are complex constant numbers and $n_{\mu \nu} \in {\mathbb N}+\{0\}$.
We assume that $D(p)=0$ at $p=\tilde{p}$.
From the Taylor expansion around $\tilde{p}$,
\be
D(\tilde{p}+\delta p) &=& \sum_{\mu,\nu,n_{\mu \nu}}  \left[ A_{\mu \nu n_{\mu \nu}} \cos(n_{\mu \nu} \tilde{p}_{\mu}) - B_{\mu \nu n_{\mu \nu}} \sin(n_{\mu \nu} \tilde{p}_{\mu}) \right] n_{\mu \nu} \delta p_{\mu} \cdot \gamma_{\nu} + O(\delta p^{2}) \nl
 &=& \sum_{\mu, \nu} \left.  \frac{\pd {\cal D}_{\nu}(p)}{\pd p_{\mu}} \right|_{p=\tilde{p}} \delta  p_{\mu} \cdot \gamma_{\nu} + O(\delta p^{2}).
\ee
If we take the continuum limit, only the $\delta p$-linear terms survive and are imposed by the following condition,
\be
\sum_{\mu} \left.  \frac{\pd {\cal D}_{\nu}(p)}{\pd p_{\mu}} \right|_{p=\tilde{p}} \delta  p_{\mu} &\ne& 0 \quad \mbox{for any} \ \nu, \label{eq:delD_cond}
\ee
because if this condition is not satisfied, $D(p)$ has no excitation modes or flat directions around $p=\tilde{p}$.
From Eq. (\ref{eq:delD_cond}) and the intermediate-value theorem, we can derive that $D(p)$ generates even doublers which are pairs of $\pm$-chirality.
However, this derivation is insufficient where $D(p)$ is not satisfied with $\gamma_{5}$hermiticity because its poles appear at not only $D(\tilde{p})=0$. 
If we assume $\gamma_{5}$hermiticity to $D(p)$, the complex constants $A_{\mu \nu n_{\mu\nu}},B_{\mu \nu n_{\mu\nu}}$ are pure imaginary.
Hence $D^{2}(p)=0$ is a necessary and sufficient condition for $D(p)=0$; namely, ${\cal D}_{\nu}(p)$ at $p=\tilde{p}$ satisfies a dispersion relation if and only if ${\cal D}_{\nu}(\tilde{p})=0$ for any $\nu$.
Therefore, $D(p)$ always generates even doublers which are $\pm$-chiral pairs, i.e. in the framework of Nielsen-Ninomiya theorem  \cite{Nielsen:1980rz}-\cite{ Nielsen:1981hk}.
On the other hand, we can obtain the Taylor expansion of $D^{(2)}_{3}(\tilde{p}+\delta p)$, which is not satisfied with $\gamma_{5}$hermiticity, around $\tilde{p}=(\pi/2,-\pi/2)$ as
\be
D^{(2)}_{3}(\tilde{p}+\delta p) &=& (-\delta p_{1} + i -1) \cdot \gamma_{1}+(\delta p_{4} - i -1) \cdot \gamma_{4} + O(\delta p^{2}).\nl \label{eq:D3taylor}
\ee  
We can see that $D^{(2)}_{3}(\tilde{p})$ is not equal to zero but the squared $D^{(2)}_{3}(\tilde{p})$ is.
In addition, the Eq. (\ref{eq:D3taylor}) does not approach the continuum fermion, $D_{{\rm c}}(p)=i \sum_{\mu} p_{\mu} \gamma_{\mu}$, in the continuum limit.
From this discussion, we can state that non-$\gamma_{5}$hermiticity affects not only odd doublers but also non-trivial doublers, which appear at $D(\tilde{p})\ne0$ and $D^{2}(\tilde{p}) = 0$.
Additionally the non-trivial doublers do not approach an appropriate continuum limit.

Finally, we prove the reflection positivity for Dirac operators defined as follows:
\be
\bar{D}^{(2)}_{1\ nm} &=& \sum_{\mu=1,4} \frac{1}{2} \left( \delta_{n+\hat{\mu},m} - \delta_{n-\hat{\mu},m} \right) \cdot \gamma_{\mu} +\frac{\kappa}{2} \left(2 \delta_{n,m}-\delta_{n+\hat{4},m} - \delta_{n-\hat{4},m} \right) \cdot \gamma_{4}, \label{eq:mod-br} \nl 
\ee
\be
\bar{D}^{(2)}_{2 \ nm} &=& \sum_{\mu=1,4} \frac{1}{2} \left( \delta_{n+\hat{\mu},m} - \delta_{n-\hat{\mu},m} \right) \cdot \gamma_{\mu} +\frac{\kappa}{2} \left(2 \delta_{n,m}-\delta_{n+\hat{1},m} - \delta_{n-\hat{1},m} \right) \cdot \gamma_{4}, \label{eq:mod-br2} \nl
\ee
\be
\bar{D}^{(2)}_{5 \ nm} &=& \sum_{\mu=1,4} \frac{1}{2} \left( \delta_{n+\hat{\mu},m} - \delta_{n-\hat{\mu},m} \right) \cdot \gamma_{\mu} +\frac{\kappa}{2} \sum_{\mu=1,4} \left(2 \delta_{n,m}-\delta_{n+\hat{\mu},m} - \delta_{n-\hat{\mu},m} \right) \cdot \gamma_{4} \label{eq:mod-br5},\nl
\ee
The reflection positivity is the unitarity condition in Euclidean space.
Although there are two kinds of reflection positivity, site-reflection and link-reflection, we prove only the link-reflection positivity from now on
\footnote{The procedure for proof of the site-reflection positivity is similar to the case of link-reflection, but we cannot prove the site-reflection positivity.}.
Now, we prove it in the only $\bar{D}^{(2)}_{1}$ case.
We can also prove the link-reflection positivity of $\bar{D}^{(2)}_{2}$ and $\bar{D}^{(2)}_{5}$ in way similar way, described below.
Here, we define an anti-linear mapping $\Theta$ acting on the fermions as follows:
\be
\Theta(\psi_{n_{1},n_{4}})=\bar{\psi}_{n_{1},1-n_{4}}\gamma_{4}, \quad
\Theta(\bar{\psi}_{n_{1},n_{4}})=\gamma_{4} \psi_{n_{1},1-n_{4}},
\ee
and on fermion bilinear,
\be
\Theta(\bar{\psi}_{n_{1},n_{4}} \Gamma \psi_{m_{1},m_{4}}) &=& \Theta(\psi_{m_{1},m_{4}}) \Gamma^{\dagger} \Theta(\bar{\psi}_{n_{1},n_{4}})  \nl
 &=& \bar{\psi}_{m_{1},1-m_{4}} \gamma_{4} \Gamma^{\dagger} \gamma_{4}\psi_{n_{1},1-n_{4}},
\ee
where $\Gamma$ is an arbitrary function depending on the $\gamma$-matrix.
Let us denote the fermions in the half-space with $n_{4} \ge 1$ by $\psi^{(+)}$ and $\bar{\psi}^{(+)}$, and in the other half-space $n_{4} \le 0$ by $\psi^{(-)}$ and $\bar{\psi}^{(-)}$.
According to the above notation, the action Eq. (\ref{eq:mod2-1}) can be written as 
\be
S=S_{+}[\psi^{(+)},\bar{\psi}^{(+)}] + S_{-}[\psi^{(-)},\bar{\psi}^{(-)}]
+S_{\mathrm{c}}[\psi^{(+)},\bar{\psi}^{(+)},\psi^{(-)},\bar{\psi}^{(-)}],
\ee 
where
\be
  S_{\mathrm{c}} &=& \frac{1-\kappa}{2} \bar{\psi}^{(-)}_{n_{1},0} \gamma_{4} \psi^{(+)}_{n_{1},1}-\frac{1+\kappa}{2} \bar{\psi}^{(+)}_{n_{1},1} \gamma_{4} \psi^{(-)}_{n_{1},0}  \nl 
 &=& -\frac{1-\kappa}{2}  \psi^{(+)\top}_{n_{1},1}  \Theta(\psi^{(+)}_{n_{1},1})^{\top} - \frac{1+\kappa}{2} \bar{\psi}^{(+)}_{n_{1},1} \Theta(\bar{\psi}^{(+)}_{n_{1},1}) ,\label{eq:Sc1}
\ee
$S_{+}$ depends only on the fermions in positive time, and related with $S_{-}$ as 
\be
\Theta(S_{+}[\psi^{(+)},\bar{\psi}^{(+)}]) &=& S^{\dagger}_{+}[\Theta(\psi^{(+)}),\Theta(\bar{\psi}^{(+)})]) \nl
 &=& S_{-}[\psi^{(-)},\bar{\psi}^{(-)}].
\ee  
For proof of the reflection positivity, we must show $\left\langle \Theta(F) F \right\rangle \ge 0$, where $F$ is an arbitrary function depending on positive time fermions, $\psi^{(+)}$ and $\bar{\psi}^{(+)}$.
$\left\langle \Theta(F) F \right\rangle$ can be written as follows:
\be
\left\langle \Theta(F) F \right\rangle &=& Z^{-1} \int D\bar{\psi}^{(+)}  D\psi^{(+)} F[\psi^{(+)},\bar{\psi}^{(+)}]  e^{-S_{+}[\psi^{(+)},\bar{\psi}^{(+)}]} \nl
 & & \cdot \int D\bar{\psi}^{(-)}  D\psi^{(-)} F^{\dagger} [\psi^{(-)},\bar{\psi}^{(-)}]  e^{-S_{-}[\psi^{(-)},\bar{\psi}^{(-)}]} \cdot e^{-S_{\mathrm{c}}}. \label{eq:positive}
\ee
From Eq. (\ref{eq:Sc1}), the expectation value $\langle \Theta(F) F \rangle$ is positive for every $-1 \le \kappa \le 1$.

We can prove the link-reflection positivity for $\bar{D}^{(2)}_{2}$ at any $\kappa$ and $\bar{D}^{(2)}_{5}$ at $-1 \le \kappa \le 1$ from the same discussion.

\section{Gross-Neveu model in two dimensions} \label{sec:aoki}
In this section, we apply the non-$\gamma_{5}$hermiticity fermion defined in Sect. \ref{sec:non-gamma5} to the 2D Gross-Neveu model and draw phase diagrams \cite{Gross:1974jv}.
The model is investigated as a toy model for QCD and can be solved exactly in the large-$N$ limit using the saddle point approximation.
In this limit, we can also obtain the parity broken phase diagram, called Aoki phase \cite{Aoki:1983qi}-\cite{Nakano:2012wa}.


Now, we apply the Dirac operator $D^{(2)}_{1}$ in Eq. (\ref{eq:mod2-1}) and study the phase diagrams.
At first sight, the fermion seems to be unsuitable for a practical calculation because it does not preserve $\gamma_{5}$hermiticity.
These analyses are simple tests for application of the fermion to a concrete model.
In this paper, we analyze two cases, (1) parity symmetry breaking and (2) chiral symmetry breaking, which is caused in the model with an imaginary chemical potential.
We also compare the phase diagrams using the fermion with those using the naive fermion.
The chiral broken phase diagrams are studied in the next section.  

Here, we define the continuum Gross-Neveu model in two dimensions as follows,
\be
S^{\mathrm{GN}}_{\mathrm{c}} = \int {d^{2}}x \left[ \bar{\psi} \left( \pd \cdot \gamma +m \right) \psi - \frac{g^{2}_{\sigma}}{2N} \left( \bar{\psi} \psi \right)^{2} -\frac{g^{2}_{\pi}}{2N} \left( i \bar{\psi} \gamma_{5} \psi  \right)^{2}  \right], \label{eq:GNc}
\ee
where we omit the flavor indices, $\bar{\psi}\psi \equiv \sum_{i=1}^{N} \bar{\psi}^{i} \psi^{i}$.
In this action, the fermions have imposed $\mathrm{U_{V}(1)}$ symmetry:
\be
\psi &\rightarrow& e^{i \theta} \ \psi,  \nl
\bar{\psi} &\rightarrow& \bar{\psi} \ e^{-i \theta}.
\ee
In massless case, the fermions preserve continuum chiral symmetry for $g^{2}_{\sigma}=g^{2}_{\pi}$ and chiral $\mathbf{Z}_{4}$ symmetry for $g^{2}_{\sigma} \ne g^{2}_{\pi}$.
Then, we introduce an auxiliary scalar field $\sigma$ and an auxiliary pseudo-scalar field $\pi$.
The partition function in the continuum theory is defined as follows, 
\be
Z_{\mathrm{c}}= \int D\psi D\bar{\psi} D\sigma D\pi \ e^{-S_{\mathrm{c,aux}}^{\mathrm{GN}}},
\ee
where 
\be
S_{\mathrm{c,aux}}^{\mathrm{GN}}=\int {d^{2}}x \left[ \bar{\psi} \left( \ \pd \cdot \gamma + m + \sigma + \pi i \gamma_{5} \  \right) \psi +  \frac{N}{2g^{2}_{\sigma}} \sigma^{2} + \frac{N}{2g^{2}_{\pi}} \pi^{2}    \right]. \label{eq:GNcaux}
\ee
Integrating out the auxiliary fields $\sigma$ and $\pi$, the action (\ref{eq:GNcaux}) recovers the former action, Eq. (\ref{eq:GNc}), from the following relations:
\be
\frac{N}{g^{2}_{\sigma}} \sigma &=& -  \bar{\psi} \psi  ,\nl
\frac{N}{g^{2}_{\pi}} \pi &=& - i  \bar{\psi} \gamma_{5} \psi.
\ee
Next, we define a lattice action from the continuum action.
We choose Dirac operator as $\bar{D}^{(2)}_{1}$, defined in Eq. (\ref{eq:mod-br}). 
Note that the Dirac operator $D^{(2)}_{1}$ is not suitable for our purpose because we study the parity broken phase diagrams.
One difference between $\bar{D}_{1}^{(2)}$ and $D^{(2)}_{1}$ is the components in the doubler suppressing terms, temporal or spatial.
$\bar{D}_{1}^{(2)}$ preserves parity symmetry but not time-reversal symmetry.
Hence, $D^{(2)}_{1}$, which has broken parity symmetry, is not fit for our purpose.
We define the lattice action of the Gross-Neveu model as follows:
\be
S_{\mathrm{lat,aux}}^{\mathrm{GN}}=\sum_{n,m}  \bar{\psi}_{n} \left[ \ \bar{D}^{(2)}_{1 \ nm} + \left( m  + \sigma_{n} + \pi_{n} i \gamma_{5}  \ \right) \delta_{n,m}  \right] \psi_{m}  + \frac{N}{2} \sum_{n} \left[ \frac{\sigma^{2}_{n}}{g^{2}_{\sigma}}  + \frac{\pi^{2}_{n}}{g^{2}_{\pi}}  \right] . \label{eq:GNlaux}
\ee
where the lower indices denote a coordinate in lattice space.
Integrating out the fermion, we can obtain an effective action $S_{\mathrm{eff}}$:
\be
Z &=& \int D\sigma_{n} D\pi_{n} e^{-NS_{\mathrm{eff}}(\sigma,\pi)},\\
S_{\mathrm{eff}}(\sigma,\pi) &=& \frac{1}{2} \sum_{n} \left[ \frac{\sigma^{2}_{n}}{g^{2}_{\sigma}}  + \frac{\pi^{2}_{n}}{g^{2}_{\pi}}  \right] -\mathrm{Tr} \  \log D,
\ee
where $D$ is defined as follows:
\be
D_{nm}=\bar{D}^{(2)}_{1 \ nm} + \left( m  + \sigma_{n} + \pi_{n} i \gamma_{5}  \ \right) \delta_{n,m}.
\ee
In the large-$N$ limit, we can integrate out the auxiliary field $\sigma$ and $\pi$ from the partition function using the saddle point approximation.
The solutions $\bar{\sigma}_{n},\bar{\pi}_{n}$ are given by the saddle point condition,
\be
\frac{\delta S_{\mathrm{eff}}(\sigma,\pi)}{\delta \sigma_{n}} = \frac{\delta S_{\mathrm{eff}}(\sigma,\pi)}{\delta \pi_{n}} = 0 \label{eq:sad_cond}.
\ee
We impose translation invariance, $\bar{\sigma}_{n}=\sigma_{0}$ and $\bar{\pi}_{n}=\pi_{0}$ for any $n$; the partition function is written as
\be
Z &=&  e^{ -V \cdot V_{\mathrm{eff}}(\sigma_{0},\pi_{0})},\\
V_{\mathrm{eff}} &=& \frac{1}{2} \left[ \frac{\sigma^{2}_{0}}{ g^{2}_{\sigma}}  + \frac{ \pi^{2}_{0}}{ g^{2}_{\pi}} \right]  - \frac{1}{V} \mathrm{Tr} \  \log D, \label{eq:Veff}
\ee
where $V$ is the volume of the system.
The last term in Eq. (\ref{eq:Veff}) is obtained using Fourier transformation: 
\be
\mathrm{Tr} \  \log D &=& V \cdot \int \frac{d^{2}k}{(2 \pi)^{2}} \ \log \left[ \left( m + \sigma_{0} \right)^{2} + \pi^{2}_{0} + H(k)  \right], \nl
&\equiv& V \cdot  \int \frac{d^{2}k}{(2 \pi)^{2}} \ {\cal E}(k),
\label{eq:trlogD}
\ee
where 
\be
H(k) = \sin^{2} k_{1} +\left[ \sin k_{4}-i \kappa(1-\cos k_{4}) \right]^{2}.
\ee
We can easily see that Eq. (\ref{eq:trlogD}) is real because 
${\cal E}(k)$ is a $k_{1}$-even function and the sum of ${\cal E}(k_{1},k_{4})$ with ${\cal E}(k_{1},-k_{4})$ is real:
\be
& & {\cal E}(k_{1},k_{4} ) + {\cal E}(k_{1},-k_{4}) \nl
&=& \log \left[ \left\{ ( m + \sigma_{0} )^{2} + \pi_{0}^{2}+ \sum_{\mu=1,4} \sin^{2} k_{\mu}-\kappa^{2}(1-\cos k_{4})^{2} \right\}^{2} \right. \nl
&& + \left. \left\{  2 \kappa \, \sin k_{4} (1- \cos k_{4})  \right\}^{2} \right].
\ee
According to the saddle point condition (\ref{eq:sad_cond}), we can obtain the following equation:
\be
\frac{\sigma_{0}}{g^{2}_{\sigma}} &=&  \int \frac{d^{2}k}{(2 \pi)^{2}} \frac{2\sigma_{m}}{\sigma_{m}^{2} + \pi^{2}_{0} +  H(k)},\\
\frac{\pi_{0}}{g^{2}_{\pi}} &=& \int \frac{d^{2}k}{(2 \pi)^{2}} \frac{2\pi_{0}}{\sigma_{m}^{2} + \pi^{2}_{0} + H(k)},
\ee
where $\sigma_{m} = m + \sigma_{0}$.
$\pi_{0}$ is an order parameter for parity symmetry breaking; hence, $\pi$ approaches zero near the critical line in the parameter space.
Near the line ,we can derive the following gap equations:
\be
\frac{\sigma_{0}}{g^{2}_{\sigma}} &=&  \int \frac{d^{2}k}{(2 \pi)^{2}} \frac{2 \sigma_{m_{\mathrm{c}}}}{\sigma_{m_{\mathrm{c}}}^{2} +  H(k)},\label{eq:cri-line1}\\
\frac{1}{g^{2}_{\pi}} &=&  \int \frac{d^{2}k}{(2 \pi)^{2}} \frac{2}{\sigma_{m_{\mathrm{c}}}^{2} + H(k)} \label{eq:cri-line2},
\ee
where $\sigma_{m_{\mathrm{c}}}=m_{\mathrm{c}}+\sigma_{0}$. 
$m_{\mathrm{c}}$ represents a critical mass depending on $g^{2}_{\sigma}$ and $g^{2}_{\pi}$.
\begin{figure}[t]
\begin{minipage}{1.0\hsize}
\begin{center}
\includegraphics[width=70mm]{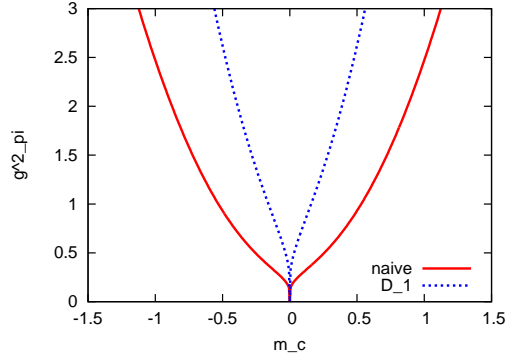}
\end{center}
\end{minipage}
\caption{The phase diagram of Aoki phase using the naive fermion (red line) and the non-$\gamma_{5}$hermiticity fermion $\bar{D}^{(2)}_{1}$ (blue line) with $g^{2}_{\sigma}=g^{2}_{\pi}/2$.
We fix the hopping parameter of the non-$\gamma_{5}$hermiticity fermion at $\kappa=1$.
The horizontal and vertical axes denote the critical mass $m_{c}$ and the squared four-fermi coupling constant $g^{2}_{\pi}$, respectively.}
\label{fig:aoki}
\end{figure}
\begin{figure}[t]
\begin{minipage}{0.5\hsize}
\begin{center}
\includegraphics[width=70mm]{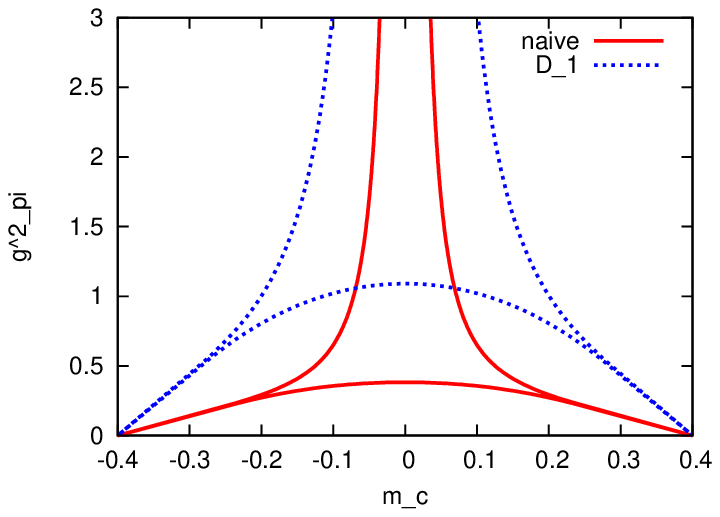}
\end{center}
\end{minipage}
\begin{minipage}{0.5\hsize}
\begin{center}
\includegraphics[width=70mm]{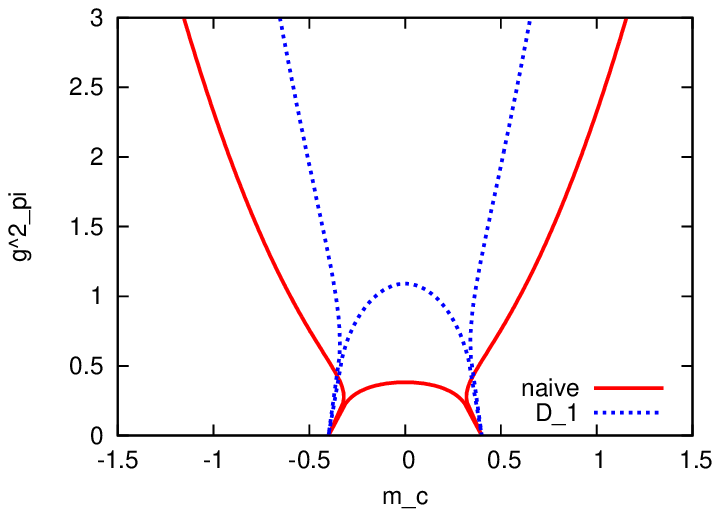}
\end{center}
\end{minipage}
\caption{The phase diagrams for the Aoki phase using the naive fermion (red line) and the non-$\gamma_{5}$hermiticity fermion $\bar{D}^{(2)}_{1}$ (blue line) adding the flavored mass term with (left) $g^{2}_{\sigma}=g^{2}_{\pi}$ and (right) $g^{2}_{\sigma}=g^{2}_{\pi}/2$.
We fix the flavored mass factor and the hopping parameter of the non-$\gamma_{5}$hermiticity fermion at $m_{f}=0.4$ and $\kappa=1$, respectively.
The horizontal and vertical axes denote the critical mass $m_{c}$ and the squared four-fermi coupling constant $g^{2}_{\pi}$ respectively.}
\label{fig:aoki-fm}
\end{figure}
At $g^{2}_{\sigma}=g^{2}_{\pi} \equiv g^{2}$, we can obtain the critical mass by dividing Eq. (\ref{eq:cri-line1}) by Eq. (\ref{eq:cri-line2}):
\be
\sigma_{0} = \sigma_{m_{\mathrm{c}}} \ \ \Rightarrow \ \ m_{c} =0 \ \ \mathrm{for \ any \ }g^{2},
\ee
and a pion mass $m^{2}_{\pi}$ on the critical line as follows:
\be
m^{2}_{\pi} &\propto& \left\langle \frac{\delta^{2} S_{\mathrm{eff}}}{\delta \pi_{0} \delta\pi_{0}}  \right\rangle \nl
 &=& V \cdot \left[ \frac{1}{g^{2}} -  \int \frac{d^{2} k}{(2 \pi)^{2}} \frac{2}{\sigma_{m_{\mathrm{c}}}^{2} + \pi^{2}_{0} + H(k)  } + \int \frac{d^{2} k}{(2 \pi)^{2}} \frac{4 \pi_{0}^{2} }{ \left( \sigma_{m_{\mathrm{c}}}^{2} + \pi^{2}_{0} + H(k) \right)^{2}  } \right] .\nl 
\ee
From (\ref{eq:cri-line2}) and the fact that $\pi_{0}$ approaches zero near the critical line, the pion mass is obtained as
\be
m^{2}_{\pi} = 0.
\ee

We also study phase diagrams with fermions adding a flavored mass term \cite{Cichy:2008gk,Creutz:2011cd}.
A flavored mass is defined as follows:
\be
m_{f}(p) &=&
\begin{cases}
m_{f} \cdot \cos p_{1} \cos p_{4} &  \mbox{for the naive fermion} \\
m_{f} \cdot \cos p_{1} &  \mbox{for the non- $\gamma_{5}$hermiticity fermion} 
\end{cases},
\ee
where $m_{f}$ on r.h.s. is constant
\footnote{We can choose some variations as a flavored mass \cite{Creutz:2011cd}.
In our definition, the doublers of non-$\gamma_{5}$hermiticity fermions defined in Eq. (\ref{eq:mod-br}) at momenta $(0,0)$ and $(0,\pi)$ have $m+m_{f}$ and $m-m_{f}$ masses  respectively.
On the other hand, the half doublers of the naive fermion at $(0,0)$ and $(\pi,\pi)$ also have $m+m_{f}$ and those at $(0,\pi)$ and $(\pi,0)$ have $m-m_{f}$.
By the definition, we can compare phase diagrams using fermions that have the same mass spectra but different numbers of doublers.}.
We add the flavored mass to the fermion mass term:
\be
m &\rightarrow& m +m_{f}(p). \nn
\ee
From this modification, the gap equations change as follows,
\be
\frac{\sigma_{0}}{g_{\sigma}^{2}} &=&  \int \frac{d^{2}k}{(2 \pi)^{2}} \frac{2 (\sigma_{m_{\mathrm{c}}}+m_{f}(k))}{(\sigma_{m_{\mathrm{c}}}+m_{f}(k))^{2} +  H(k)},\label{eq:gap-fm}\\
\frac{1}{g_{\pi}^{2}} &=& \int \frac{d^{2}k}{(2 \pi)^{2}} \frac{2}{(\sigma_{m_{\mathrm{c}}}+m_{f}(k))^{2} + H(k)} \label{eq:gap-fm2}.
\ee

We present the phase diagram without the flavored mass in Fig. \ref{fig:aoki} and with the flavored mass in Fig. \ref{fig:aoki-fm}
\footnote{In the phase diagrams using the fermions plus the flavored mass, there is 1st-order phase transition at the bottom of the diagrams enclosing the critical line.
Because of that, we cannot use the gap equation, Eqs. (\ref{eq:gap-fm}) and (\ref{eq:gap-fm2}), in this area. 
However, we ask leave not to correct this, to emphasize that we can obtain solutions for the gap equations.}.
We fix the hopping parameter at $\kappa=1$.
The four-fermi couplings in the action are related as $g^{2}_{\sigma}=g^{2}_{\pi}$ in Fig. \ref{fig:aoki}.
In Fig. \ref{fig:aoki-fm}, we fix the hopping parameter and the flavored mass factor at $\kappa=1$, and $m_{f}=0.4$ respectively, and the four-fermi couplings are related as (left) $g^{2}_{\sigma}=g^{2}_{\pi}$ and (right) $g^{2}_{\sigma}=g^{2}_{\pi}/2$.
We clearly see that the phase diagrams using the non-$\gamma_{5}$hermiticity fermion have a very similar structure to those using the naive fermion.
The critical mass and the four-fermi couplings are real numbers, despite using a fermion without $\gamma_{5}$hermiticity.


\section{Gross-Neveu model with imaginary chemical potential} \label{sec:ichem}
In this section, we focus on the 2D Gross-Neveu model with an imaginary chemical potential and study the chiral broken phase diagrams \cite{Hasenfratz:1983ba,Misumi:2012uu}.
The outline for obtaining the chiral broken phase diagram is the same as in Sec. \ref{sec:aoki}.

We define the 2D Gross-Neveu model adding an imaginary chemical potential term as follows:
\be
S^{\mathrm{GN}}_{\mathrm{c}} &=& \int {d^{2}}x \left[ \bar{\psi} \left( \pd \cdot \gamma +m \right) \psi \right. \nl
 & & \left. - \frac{g^{2}_{\sigma}}{2N} \left( \bar{\psi} \psi \right)^{2} - \frac{g^{2}_{\pi}}{2N} \left( i \bar{\psi} \gamma_{4} \psi  \right)^{2}  + i\mu \ \bar{\psi} \gamma_{4} \psi \right],\nl \label{eq:GN-chem}
\ee
where $\mu$ is a chemical potential.
Note that the third term in the action is different from Eq. (\ref{eq:GNc}), replacing $\gamma_{5}$ with $\gamma_{4}$. 
This action is imposed $\mathrm{U_{V}(1)}$ symmetry.
In addition, the action preserves chiral $\mathbf{Z}_{4}$ symmetry for $m=0$ and any $g^{2}_{\pi}$.

Introducing an auxiliary scalar field $\sigma$ and an auxiliary vector field $\pi_{4}$, we rewrite the action as follows:
\be
S^{\mathrm{GN}}_{\mathrm{c}} &=& \int {d^{2}}x \left[ \bar{\psi} \left( \pd \cdot \gamma  + \sigma + \pi_{4} i \gamma_{4} \right) \psi \right. \nl
 & & \left. + \frac{N}{2g^{2}_{\sigma}} \left( \sigma-m \right)^{2} + \frac{N}{2g^{2}_{\pi}} \left( \pi_{4}-\mu  \right)^{2} \right]. \nl \label{eq:GN-chem-aux}
\ee
where
\be
\frac{N}{g^{2}_{\sigma}} (\sigma-m) &=& -\bar{\psi}\psi,\\
\frac{N}{g^{2}_{\pi}} (\pi_{4}-\mu) &=& -i \bar{\psi}\gamma_{4}\psi.
\ee
In this analysis, we adopt $D^{(2)}_{1}$, defined in Eq. (\ref{eq:mod2-1}), as a lattice fermion.
We cannot apply $\bar{D}^{(2)}_{1}$, defined in Eq. (\ref{eq:mod-br}), because its determinant is always a complex number.
In other words, the coupling constants in the chiral phase diagrams are always complex numbers.
We discuss this issue in the next section.

We discretize space-time and write down a lattice action as follows: 
\be
S_{\mathrm{lat,aux}}^{\mathrm{GN}} &=&\sum_{n,m}  \bar{\psi}_{n} \left[ \ D^{(2)}_{nm} + \left(  \sigma_{n} + \pi_{4 n} i \gamma_{4}  \ \right) \delta_{n,m}  \right] \psi_{m} \nl
& & + \frac{N}{2} \sum_{n} \left[ \frac{1}{g^{2}_{\sigma}} (\sigma_{n}-m)^{2} + \frac{1}{g^{2}_{\pi}} (\pi_{4 n}-\mu)^{2}  \right] . \label{eq:GNlchem}
\ee
Integrating fermions in the action, an effective action is obtained as follows:
\be
Z &=& \int D\sigma_{n} D\pi_{4 n} e^{-NS_{\mathrm{eff}}(\sigma,\pi_{4})},\\
S_{\mathrm{eff}}(\sigma,\pi_{4}) &=& \frac{1}{2} \sum_{n} \left[ \frac{1}{g^{2}_{\sigma}}(\sigma_{n}-m )^{2} + \frac{1}{g^{2}_{\pi}} (\pi_{4 n}-\mu)^{2}  \right] -\mathrm{Tr} \  \log D,
\ee
where
\be
\mathrm{Tr} \,  \log D &=& V \cdot \int \frac{d^{2}k}{(2 \pi)^{2}} \ \log \left[  \sigma_{n} ^{2}  + \tilde{H}(k)  \right], 
\ee
and 
\be
\tilde{H}(k) =\left[ \sin k_{1}-i \kappa(1-\cos k_{1}) \right]^{2}
              +(\sin k_{4}+\pi_{4 n})^{2}.
\ee
\begin{figure}[t]
\begin{minipage}{0.5\hsize}
\begin{center}
\includegraphics[width=70mm]{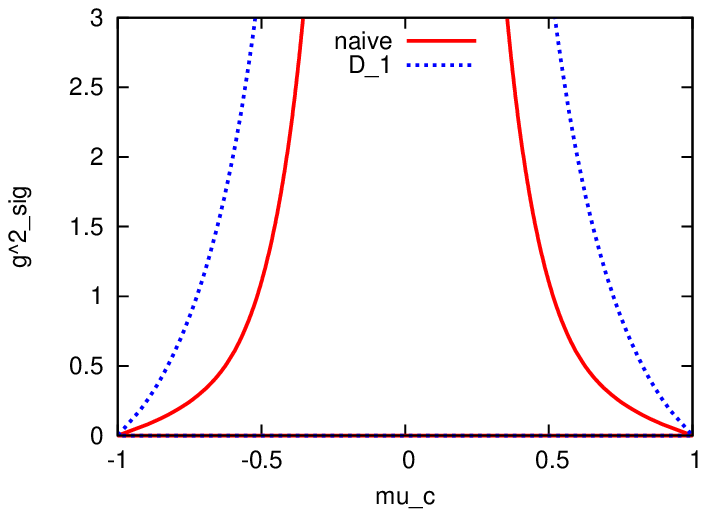}
\end{center}
\end{minipage}
\begin{minipage}{0.5\hsize}
\begin{center}
\includegraphics[width=70mm]{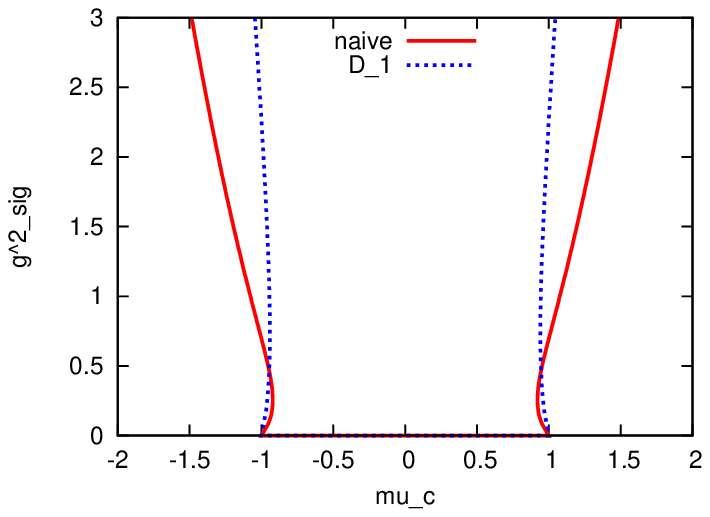}
\end{center}
\end{minipage}
\caption{The chiral broken phase diagrams using the naive fermion (red line) and the non-$\gamma_{5}$hermiticity fermion $D^{(2)}_{1}$ (blue line) with (left) $g^{2}_{\sigma}=g^{2}_{\pi}$ and (right) $g^{2}_{\sigma}=2g^{2}_{\pi}$.
We choose the hopping parameter in the non-$\gamma_{5}$hermiticity fermion to be $\kappa=2$.
The horizontal and vertical axes denote the critical chemical potential and the squared four-fermi coupling constant $g^{2}_{\sigma}$.}
\label{fig:im_chem}
\end{figure}
We can integrate out the auxiliary fields $\sigma$ and $\pi_{4}$ in the large $N$ limit, and solutions are obtained from the saddle point approximation:
\be
\frac{\delta S_{\mathrm{eff}}(\sigma,\pi_{4})}{\delta \sigma_{n}} = \frac{\delta S_{\mathrm{eff}}(\sigma,\pi_{4})}{\delta \pi_{4n}} = 0 \label{eq:sad_cond-chem}.
\ee
Imposing translation invariance on the solutions $\bar{\sigma}_{n}=\sigma_{0}$ and $\bar{\pi}_{4n}=\pi_{40}$, we can derive the following gap equations,
\be
\frac{\sigma_{0}-m}{g^{2}_{\sigma}} &=&  \int \frac{d^{2}k}{(2 \pi)^{2}} \frac{2\sigma_{0}}{\sigma_{0}^{2} + \tilde{H}(k)},\label{eq:cri-line-chem1} \\
\frac{\pi_{40}-\mu}{g^{2}_{\pi}} &=& \int \frac{d^{2}k}{(2 \pi)^{2}} \frac{2(\pi_{40}+\sin k_{4})}{\sigma_{0}^{2} + \tilde{H}(k)},\label{eq:cri-line-chem2}
\ee
Now we fix the fermion mass at $m=0$ and draw the chiral broken phase diagrams.
The auxiliary field $\sigma$ approaches zero near the critical line because $\sigma$ is an order parameter for chiral symmetry breaking.
Hence, gap equations for the chiral broken phase diagrams are derived as follows:
\be
\frac{1}{g^{2}_{\sigma}} &=&  \int \frac{d^{2}k}{(2 \pi)^{2}} \frac{2}{ \tilde{H}(k)},\label{eq:gap-chem1} \\
\frac{\pi_{40}-\mu_{c}}{g^{2}_{\pi}} &=& \int \frac{d^{2}k}{(2 \pi)^{2}} \frac{2(\pi_{40}+\sin k_{4})}{ \tilde{H}(k)},
\ee
where $\mu_{c}$ is a critical chemical potential.
At $g^{2}_{\sigma}=g^{2}_{\pi} \equiv g^{2}$, we can also obtain a meson mass $m^{2}_{\sigma}$ on the critical line from the following equation:
\be
m^{2}_{\sigma} &\propto& \left\langle \frac{\delta^{2} S_{\mathrm{eff}}}{\delta \sigma_{0} \delta \sigma_{0}}  \right\rangle \nl
 &=& V \cdot \left[ \frac{1}{g^{2}} -  \int \frac{d^{2} k}{(2 \pi)^{2}} \frac{2}{\sigma_{0}^{2}  + \tilde{H}(k)  } +  \int \frac{d^{2} k}{(2 \pi)^{2}} \frac{4 \sigma^{2}_{0}}{ (  \sigma_{0}^{2}  + \tilde{H}(k) )^{2}  } \right] .\nl  
\ee
From Eq. (\ref{eq:gap-chem1}) and the fact that the $\sigma_{0}$ approach zero on the critical line, we can obtain the meson mass on the critical line,
\be
m^{2}_{\sigma} = 0.
\ee

The chiral broken phase diagrams with the naive fermion and the non-$\gamma_{5}$hermiticity fermion $\bar{D}^{(2)}_{1}$ are presented in Fig.\ref{fig:im_chem}.
In this figure, we set the hopping parameter at $\kappa=2$, and the four-fermi coupling constants $g^{2}_{\sigma}$ and $g^{2}_{\pi}$ are related as (left) $g^{2}_{\sigma}=g^{2}_{\pi}$ and (right) $g^{2}_{\sigma}=2g^{2}_{\pi}$.
We can see that the phase diagram using the naive fermion has a qualitatively very similar structure to that using the non-$\gamma_{5}$hermiticity fermion.
As with the non-chemical potential case in Sect. \ref{sec:aoki}, all of the couping constants in the phase diagrams are real numbers.

\section{Reality condition} \label{sec:real-cond}
\begin{figure}[t] 
\begin{minipage}{0.5\hsize}
\begin{center}
\includegraphics[width=70mm]{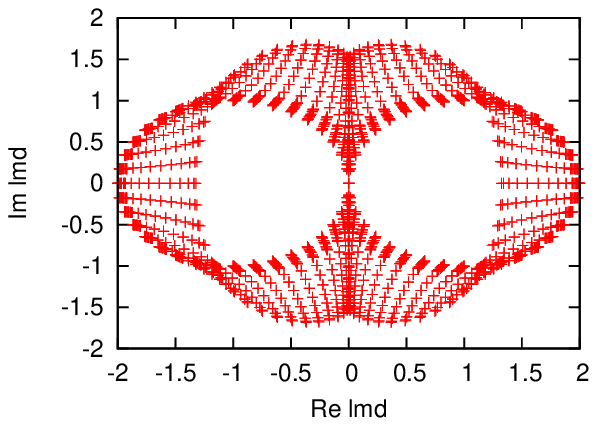}
\end{center}
\end{minipage}
\begin{minipage}{0.5\hsize}
\begin{center}
\includegraphics[width=70mm]{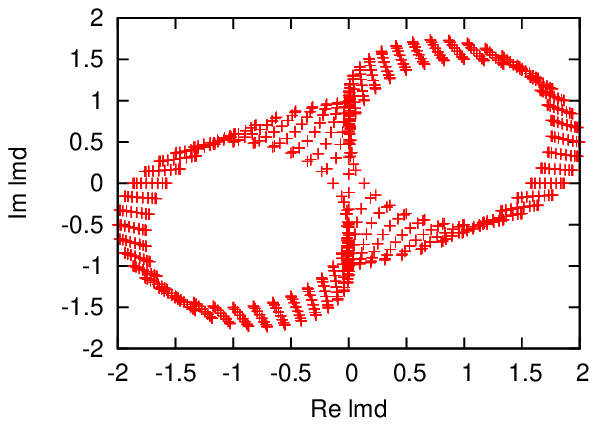}
\end{center}
\end{minipage}
\caption{The eigenvalue distribution of the Dirac operators with an imaginary chemical potential: (left) $D^{(2)}_{c}$ and (right) $\bar{D}^{(2)}_{c}$.  We set the hopping parameter and the chemical potential at $\kappa=1.0$ and $\mu=0.5$, respectively. The lattice size is $36 \times 36$. The blue circle points denote eigenvalues of momenta $p=(0,0),$ $(0,\pi)$, $(\pi,0)$, and $(\pi,\pi)$.}
\label{fig:eigenvalue-chem}
\end{figure}

In Sects. \ref{sec:aoki} and \ref{sec:ichem}, we investigated the parity and chiral phase diagrams for the Gross-Neveu model using the non-$\gamma_{5}$hermiticity fermion.
The results showed not only that the phase structure using the non-$\gamma_{5}$hermiticity fermion has a qualitatively similar structure to the naive fermion, but also that the coupling constants in the phase diagram are real numbers despite breaking $\gamma_{5}$hermiticity. 
In this section, we discuss why the coupling constants in the phase diagrams drawn in Sects. \ref{sec:aoki} and \ref{sec:ichem} are real numbers:


To see the reason, we compare two fermions adding an imaginary chemical potential:
\be
D^{(2)}_{c}(p)&=& \sum_{\mu=1,4} i \ \sin p_{\mu} \cdot \gamma_{\mu}+ \kappa(1- \cos p_{1}) \cdot \gamma_{1}+ i \mu \cdot \gamma_{4} \equiv \sum_{\mu=1,4} f_{\mu}(p) \cdot \gamma_{\mu}.\label{eq:Dim1st} \nl \\
\bar{D}^{(2)}_{c}(p)&=&  \sum_{\mu=1,4} i \ \sin p_{\mu} \cdot \gamma_{\mu}  + \kappa(1-\cos p_{4}) \cdot \gamma_{4}+i \mu \cdot \gamma_{4} \equiv  \sum_{\mu=1,4} \bar{f}_{\mu}(p) \cdot \gamma_{\mu} , \label{eq:Dim4th} \nl
\ee
The former has a $\gamma_{5}$hermiticity breaking term with a temporal index, and the latter has one with a spatial index.
Firstly, we present the eigenvalue distribution of these fermions in Fig.\ref{fig:eigenvalue-chem}.
Because a determinant of the Dirac operator is obtained by the product of all of the eigenvalues, the eigenvalues of the Dirac operator must be complex conjugate pairs for a real determinant. 
Figure \ref{fig:eigenvalue-chem} shows that all the eigenvalues obtained from $D^{(2)}_{c}$ have complex conjugate pairs, but the eigenvalues of $\bar{D}^{(2)}_{c}$ do not. 
The Dirac operators seem to have very similar forms, but only $D^{(2)}_{c}$  preserves hermiticity, which indicates a real determinant.

To investigate in more detail, we obtain the products of the eigenvalues of the Dirac operators, namely determinants.
The determinants are obtained as follows:
\be
\det D^{(2)}_{c} &=& \prod_{p} \sum_{\mu=1,4} (-1) \cdot f_{\mu}^{2}(p), \label{eq:detDim1st} \\ 
\det \bar{D}^{(2)}_{c} &=& \prod_{p} \sum_{\mu=1,4}(-1) \cdot \bar{f}_{\mu}^{2}(p), \label{eq:detDim4th}
\ee
where $f_{\mu}(p)$ and $\bar{f}_{\mu}(p)$ are defined in Eqs. (\ref{eq:Dim1st}) and (\ref{eq:Dim4th}).
A determinant of a Dirac operator that preserves $\gamma_{5}$hermiticity is a real number because $\sum_{\mu=1,4}f_{\mu}^{2}(p)$ for any momentum is always real. 
By contrast, a determinant of fermion without $\gamma_{5}$hermiticity is, in general, a complex number.
However, if an arbitrary Dirac operator $D(p)= \sum_{\mu} {\cal F}_{\mu}(p) \cdot \gamma_{\mu}$ satisfies the following condition, we can obtain its Dirac determinant, which is real:
\be
\sum_{\mu=1,4} {\cal F}^{2}_{\mu}(p) = \sum_{\mu=1,4} {\cal F}^{2 \, *}_{\mu}(\tilde{p}) \quad \mbox{for any} \ p, \label{eq:real_cond}
\ee
where ${*}$ denotes complex conjugate and $\tilde{p}$ is a momentum defined by the condition.
The fermion $D_{c}^{(2)}$ defined in Eq. (\ref{eq:Dim1st}) satisfies the condition at $\tilde{p}=(-p_{1},p_{4})$; hence, the determinant of $D^{(2)}_{c}$ is a real number.
On the other hand, we cannot obtain the determinant of $\bar{D}^{(2)}_{c}$ as a real number because $\bar{D}_{c}^{(2)}$ cannot satisfy the condition for any $\tilde{p}$.
This fact suggests that although $\mathrm{det} \, D_{c}^{(2)}(p_{1},p_{4})$ which is a part of a determinant of $D^{(2)}_{c}$ is a complex number, it is real-valued by the product with $\mathrm{det} \, D_{c}^{(2)}(-p_{1},p_{4})$.
In the $\bar{D}^{(2)}_{c}$ case, however, $\mathrm{det} \, \bar{D}^{(2)}_{c}(p)$ cannot be real-valued by any momentum mode. 
Therefore, Eq. (\ref{eq:real_cond}) guarantees reality for the fermions, at least, in free theory and Yukawa theory.

\section{Summary and discussion} \label{sec:sum}
We have constructed fermions without $\gamma_{5}$hermiticity (non-$\gamma_{5}$hermiticity fermions) based on the minimal doubling fermion in two dimensions and investigated symmetries and properties of the fermions.
The fermions preserve translation invariance, chiral symmetry and locality but broken cubic symmetry and some discrete symmetries.
To investigate the model application possibilities, we have studied the eigenvalue distribution and the number of poles for the fermions.
The eigenvalues of $D^{(2)}_{1}$, defined in Eq. (\ref{eq:mod2-1}), are distributed along the imaginary axis in the continuum limit.
However, the eigenvalues of the other operators, defined in Eqs. (\ref{eq:mod2-2})-(\ref{eq:mod2-5}), are distributed in the entire plane in the limit.
We can also see that the fermions have more than four or odd poles in general.
In the $D^{(2)}_{1}$ case, only two poles appear at $\kappa \ge 1$. 
We have also stated that non-$\gamma_{5}$hermiticity affects not only odd doublers but also non-trivial doublers that appear at $D(\tilde{p})\ne0$ and $D^{2}(\tilde{p}) = 0$.
Also, the non-trivial doublers do not approach an appropriate continuum limit.
In addition, we have proved the link-reflection positivity in the $\bar{D}^{(2)}_{1}$ case.
From the proof, the hopping parameter is restricted to $|\kappa| \le 1$.

As simple tests for application to a concrete model, we studied the parity broken phase diagram, called Aoki phase, for the 2D Gross-Neveu model using the non-$\gamma_{5}$hermiticity fermion.
We also studied the chiral broken phase diagram for the Gross-Neveu model, adding an imaginary chemical potential in the massless case. 
Both phase diagrams are qualitatively very similar to those using the naive fermion.
All of the coupling constants in the analyses were obtained as real numbers, despite using a fermion without $\gamma_{5}$hermiticity.

We have discussed the reason for the reality for the Gross-Neveu model using the non-$\gamma_{5}$hermiticity fermion, $D^{(2)}_{1}$. 
To understand this, we investigated the eigenvalue distribution of $D^{(2)}_{c}$ and $\bar{D}^{(2)}_{c}$, defined in Eqs. (\ref{eq:Dim1st}) and (\ref{eq:Dim4th}).
We can see that the eigenvalues of $D^{(2)}_{c}$ are complex conjugate pairs, but the eigenvalues of $\bar{D}^{(2)}_{c}$ are not.
Hence, the determinant of $D^{(2)}_{c}$ is a real number but $\bar{D}^{(2)}_{c}$ is not.
Although a determinant of $ D^{(2)}_{c}(p1,p2)$ is obtained as a complex number, it is real-valued by the product with a determinant of $ D^{(2)}_{c}(-p1,p4)$.
Therefore all the coupling constants in the theory using the non-$\gamma_{5}$hermiticity fermion $D^{(2)}_{1}$ are real numbers, and the theory preserves the hermiticity, despite broken the $\gamma_{5}$hermiticity.

Incidentally, the non-$\gamma_{5}$hermiticity fermions $\bar{D}^{(2)}_{1}$ defined in Eq. (\ref{eq:mod-br}), replacing the hopping parameter $\kappa$ with a chemical potential $\mu$, seem to be fermions with a momentum-dependent chemical potential added.
In this interpretation, half of the doublers that appear at momenta $(0,0),(\pi,0)$ in the $\bar{D}^{(2)}_{1}$ case have zero-chemical potential; for the others, it is  $2\mu$. 
As is well known, the continuum fermion with a chemical potential has eigenvalues distributed entirely, like $D_{2}^{(2)}$, in the ${\rm Re}\lambda$-${\rm Im}\lambda$ plane.
In the $\bar{D}^{(2)}_{1}$ case, there are undistributed spaces, which is a lattice artifact in the plane.
Thanks to the lattice artifact, we might be able to obtain observables without regard to the mass or boundary condition at finite lattice spacing, at least, in free theory or Yukawa theory.

We will investigate gauge theory using a non-$\gamma_{5}$hermiticity fermion and higher-dimensional extension in future work.

\subsection*{Acknowledgment} \mbox{}\indent
S.K. thanks T. Misumi for helpful discussions and comments.
This work was partially supported by the Research Center for Measurement in Advanced Science of Rikkyo University.

\end{document}